\documentclass[a4paper]{article}
\usepackage{spconf,amsmath,graphicx}
\usepackage{multirow}
\usepackage{stfloats}
\usepackage{color}
\usepackage{booktabs}

\usepackage[colorlinks,linkcolor=blue]{hyperref}
\usepackage{setspace}

\title{Exploring  Voice Conversion based Data Augmentation in Text-Dependent Speaker Verification}
\name{Xiaoyi Qin$^{1,2}$, Yaogen Yang$^{2,3}$, Lin Yang$^{4}$, Xuyang Wang$^{4}$, Junjie Wang$^{4}$, Ming Li$^{1,2}$\sthanks{This research is funded in part by the National Natural Science Foundation of China (61773413), Key Research and Development Program of Jiangsu Province (BE2019054), Six talent peaks project in Jiangsu Province (JY-074), Science and Technology Program of Guangzhou City (201903010040,202007030011) and Lenovo.}}
  
  \address{$^{1}$School of Computer Science, Wuhan University, Wuhan, China  \\
      $^{2}$Data Science Research Center, Duke Kunshan University, Kunshan, Chain\\
      $^{3}$School of Electronics and Information Technology, Sun Yat-sen University, Guang Zhou, China \\
      $^{4}$AI Lab of Lenovo Research, Beijing, China\\
      \url{ming.li369@dukekunshan.edu.cn}}

\begin{document}
%
\maketitle
\begin{abstract}
In this paper, we focus on improving the performance of the text-dependent speaker verification system in the scenario of limited training data. The speaker verification system deep learning based text-dependent generally needs a large scale text-dependent training data set which could be labor and cost expensive, especially for customized new wake-up words. In recent studies, voice conversion systems that can generate high quality synthesized speech of seen and unseen speakers have been proposed.  Inspired by those works, we adopt two different voice conversion methods as well as the very simple re-sampling approach to generate new text-dependent speech samples for data augmentation purposes. Experimental results show that the proposed method significantly improves the Equal Error Rare performance from 6.51\% to 4.51\% in the scenario of limited training data.

\end{abstract}


\section{Introduction}
Speaker verification technology aims to determine whether the testing utterance is indeed spoken by enrollment speaker. In recent years, x-vectors\cite{snyder_x-vectors} and their subsequent\cite{ftdnn} demonstrate provide state-of-the-art results in speaker verification field. The backbone architectures of TDNN\cite{snyder_x-vectors}, ResNet\cite{cai_exploring_2018}, and their variants\cite{ftdnn} are usually adopted for the front-end feature extraction.

On the other hand, the research works of deep learning based speaker verification also enjoys many publicly open and free speech databases, e.g., AISHELL2 \cite{aishell2_2018}, Librispeech \cite{librispeech}, Voxceleb1\&2  \cite{nagrani_voxceleb:_2017,chung_voxceleb2_2018} in the text-independent field, RSR2015 \cite{rir2015}, HIMIA \cite{himia_dataset},  MobvoiHotwords in the text-dependent field, etc. \cite{google_2016,google_2018} achieved a good performance in text-dependent speaker verification task under the  a large amount of text-dependent data scenario. However, it is both labor expensive and time consuming to collect the database.  With the rise of smart home and Internet of Things applications, there are great demands for text-dependent speaker verification, with customized wake-up words. It is almost impossible to collect the corresponding text-dependent speech data for each smart assistant's customized wake-up word.

In recent studies, the speech signals generated by the multi-speaker Text-to-Speech (TTS) and one-to-many or many-to-many voice conversion (VC) systems are getting harder to be distinguished between real-person voice and synthesized voice. So, it is natural to adopt TTS or VC as a data augmentation strategy for speaker verification under the limited training data scenario. The multi-speaker TTS system could create a large amount of speech data from multiple target speakers with different lexical contents. However, in the context of text-dependent cases, since the input text is the same, the synthesized speech data are very similar even for different target speakers. Moreover, different from multi-speaker TTS, the VC system can generate data with various kinds of styles all with the same text-dependent content. Therefore, the VC approaches are more appropriate than TTS as the data augmentation method for text-dependent speaker verification.

This paper aims to improve the text-dependent speaker verification system's performance with a limited number of speakers and training data. 
\begin{itemize}
	\item Limited training data for each speaker. The number of text-dependent utterances of each speaker is less than 10 or less.
	\item Limited speakers for training. The number of speakers is less than 500 or less.
\end{itemize}
Targeting the aforementioned scenarios, we propose to train a voice conversion model with limited existing text-dependent data to generate more new text-dependent data. The proposed many-to-many voice conversion system using the conditional Seq-to-Seq neural network framework with dual speaker embedding as well as the phoneme posterior probability(PPP)\cite{ppp_feature} with target speaker embedding to Mel-spectrogram system\cite{ppp2mel} is employed as our data augmentation systems. Furthermore, in the limited speaker number case, we adopt the pitch shift(speed perturbation with re-sampling) strategy to augment more speakers. Besides, we also attempt to use the out of set unseen speakers' embeddings to generate the text-dependent data from out of set speakers. Finally, the ResNet34-GSP\cite{cai_exploring_2018} model is adopted as the speaker verification system to measure the strategy.

The paper is organized as follows. Section 2 describes the related work about voice conversion and speaker verification we adopted in this paper. The proposed methods and strategies are described in section3. Section 4 shows the experimental setup and results. Finally, the conclusion is provided in section 5. 

\begin{figure}[t]
  \centering
  \includegraphics[width=\linewidth]{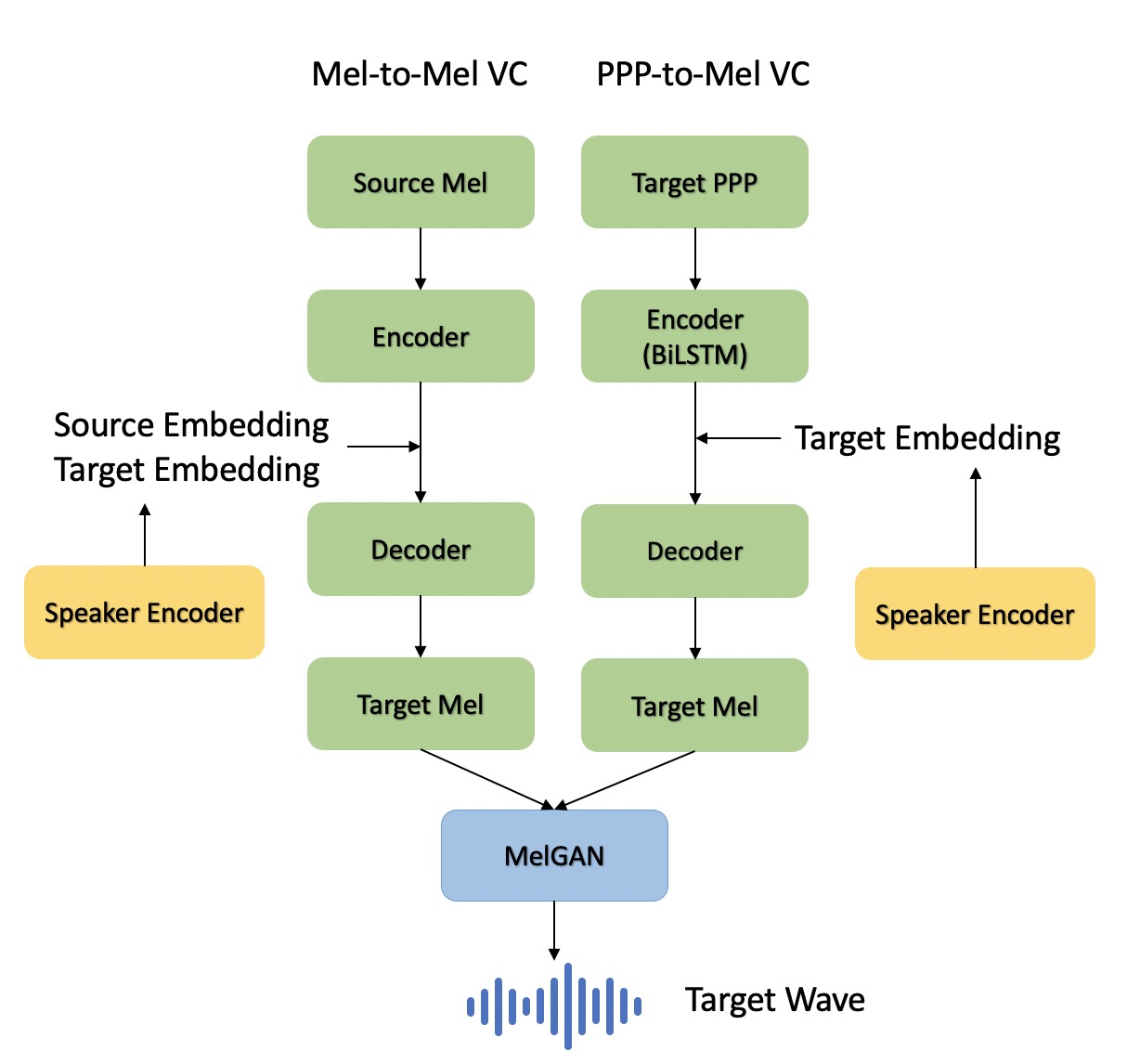}
  \caption{The architectures of two voice conversion models}
  \label{fig:vc_arc}
\end{figure} 
 
\section{Related Works}

\subsection{Speaker Verification System}

In this paper, we adopt the same structure as \cite{cai_exploring_2018}. The network structure contains three main components: a front-end pattern extractor, an encoder layer, and a back-end classifier. The ResNet34\cite{He2016Deep} structure is employed as the front-end pattern extractor, which learns a frame-level representation from the input acoustic feature. The global statistic pooling (GSP) layer, which computes the mean and standard deviation of the output feature maps, can project the variable length input to the fixed-length vector. The output of a fully connected layer followed after the pooling layer is adopted as the speaker embedding layer. The ArcFace\cite{arcface} (s=32,m=0.2) which could increase intra-speaker distances while ensuring inter-speaker compactness is used as a classifier . The detailed configuration of the neural network is the same with\cite{ffsvc20}. The cosine similarity serves as the back-end scoring method.

\subsection{Voice Conversion System}
\subsubsection{Many-to-Many VC System}


Firstly, we introduce a many-to-many voice conversion model using the conditional sequence-to-sequence neural network framework with dual speaker embedding. The model is trained on many different source-target speaker pairs, which requires the speaker embeddings from both the source speaker and the target speaker as the auxiliary inputs. To improve speaker similarity between reference speech and converted speech, we use a feedback constraint mechanism\cite{Cai2020}, which adds an auxiliary speaker identity loss in the network. This model is named as the Mel-to-Mel VC system caused the model adopts the mapping of source speaker Mel-spectrogram to target speaker Mel-spectrogram.


 \begin{figure}[htbp]
  \centering
  \includegraphics[width=6cm,height=6cm]{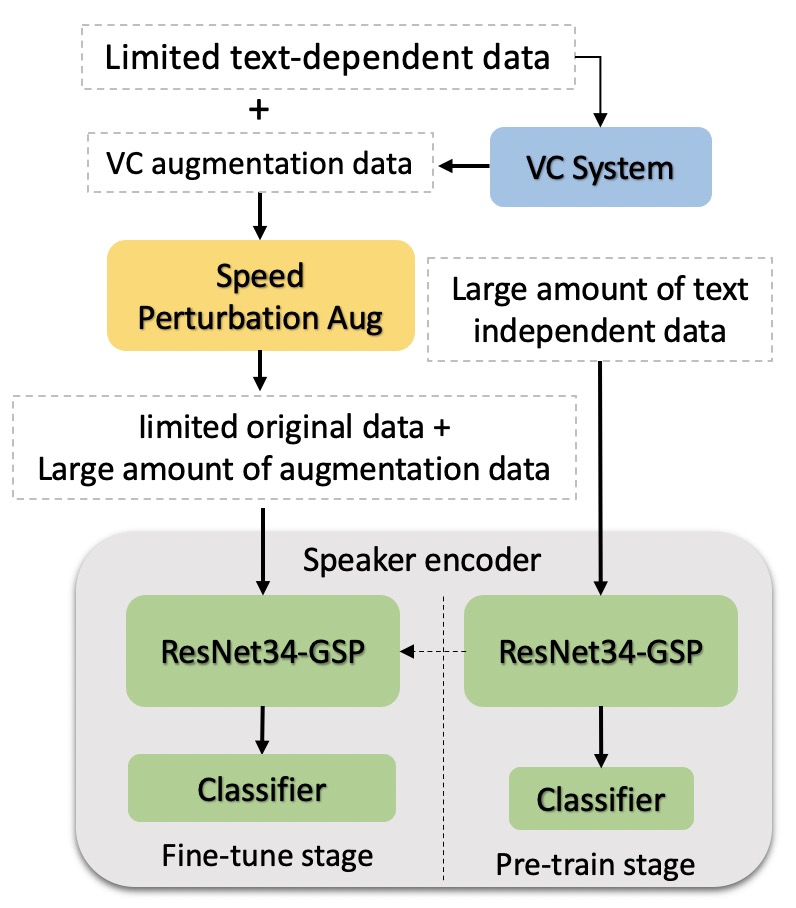}
  \caption{The pipeline of Data Augmentation based on Voice Conversion in Text-Dependent Speaker Verification.}
  \label{fig:pipline}
\end{figure}

\subsubsection{PPP-to-Mel VC System}

Besides, we also introduced another VC conversion system. The model is proposed in \cite{ppp2mel}. Firstly, we use a DNN based ASR acoustic model, trained on the AISHELL-2 database, to obtain the target speaker phoneme posterior probabilities(PPP) feature as the voice conversion model's input. The model's output is still the Mel-spectrogram feature. Similarly, the system is named the PPP-to-Mel VC system in this paper. The PPP-to-Mel VC system architecture is similar to the Mel-to-Mel VC system expect that there is no feedback constraint.  Besides, since the system's input is the target speaker feature rather than the source speaker feature, the input PPP feature is selected randomly from the limited training data.

 Fig.\ref{fig:vc_arc} shows the architectures of two conversion system. The component of the speaker encoder is adopted  the ResNet34-GSP model mentioned above. The vocoder MelGAN\cite{melgan} can be used to reconstruct the time-domain waveform from predicted Mel-spectrogram.

\section{Methodology}

In this section, the VC data augmentation strategies and the speed perturbation method will be detailed descriptions. The pipeline of our proposed data augmentation strategy is shown as Fig.\ref{fig:pipline}. Those methods are all focused on the limited text-dependent data scenario. In this experiment, we adopt the HIMIA database with 340 speakers\cite{himia_dataset} as the limited text-dependent training data.


 In this paper, 9 utterances of each speaker in the HIMIA database are randomly chosen as limited text-dependent data scenario to train the baseline system. Therefore, only 3060 utterances (total have 340*9=3060 utterances) were used to train the VC conversion and fine-tune speaker verification models. The close-talk text-dependent data of the FFSVC20\cite{ffsvc20} was chosen as test data. The trial file can be download from \href{https://github.com/qinxiaoyi/VCaug\_ASV}{trial\_file} 
 
\subsection{Pre-training and Fine-tuning}

According to our previous work\cite{xiaoyi_farfield,himia_dataset}, fine-tuning is an effective transfer learning approach to improve the speaker verification system performance in limited training data scenario. In this paper, we pre-trained the deep speaker verification network with a large scale text-independent mix-dataset. There are total 3742 speakers in pre-training data, including AISHELL-2\cite{aishell2_2018}, SLR68\footnote{https://openslr.org/68/} and SLR62\footnote{https://openslr.org/62/} from openslr.org. The three databases mentioned above are also treated as out of set unseen speaker data for the VC augmentation system. The model was trained for 200 epochs in the pre-training stage, with an initial learning rate of 0.1. The network was optimized by stochastic gradient descent(SGD). All weights in the network remain trainable with an initial learning rate of 0.01  during the fine-tuning stage.

\subsection{Data augmentation based on the VC system} 
%

\begin{table*}[tp]
  \caption{The performance of the text dependent speaker verification systems under different data augmentation methods. the $9 utt$ denotes the limited training data scenario, each speaker only has 9 utterances; the VC AUG$_{in}$ and the VC AUG$_{out}$ denotes the voice conversion data from in-set and out-of-set speakers respectively;  the Pitch shift AUG denotes the SoX  $speed$ function based pitch shift augmentation method.}
  \label{tab:results}
  \centering
  \begin{tabular}[c]{llllll}
    \toprule
     \textbf{Model} & \textbf{Training data} &  \textbf{Spk / Utt Num.}  &\textbf{EER[\%]}  & \textbf{mDCF}$_{0.1}$ \\
      
      \midrule
   \multirow{2}*{\shortstack{Pre-train \\ model}} & \multirow{2}*{AISHELL2 +SLR62 +SLR68} & \multirow{2}*{3472 / }  \multirow{2}*{518864} & \multirow{2}*{6.51} & \multirow{2}*{0.265}  \\
   & & & & \\
   \midrule
   \multirow{7}*{\shortstack{Fine-tune\\model}} & 9 utt (baseline system) & 340 / 3060 & 7.63 & 0.331\\
       & \quad + Pitch shift AUG & 1020 / 9180 & 5.76 & 0.248\\
    & \quad + VC AUG$_{in}$(Mel-to-Mel)& 340 / 26160 & 6.36 & 0.304 \\
        & \quad  + VC AUG$_{in}$(PPP-to-Mel) & 340 / 29089 & 5.16 & 0.249\\
    & \quad + VC AUG$_{out}$(Mel-to-Mel)& 3210 / 48890 & 6.08 & 0.295 \\
    & \quad + VC AUG$_{in}$(Mel-to-Mel) + Pitch shift AUG & 1020 / 76978& 5.19& 0.241 \\
    & \quad  + VC AUG$_{in}$(PPP-to-Mel) + Pitch shift AUG & 1020 / 87267& \textbf{4.51}& \textbf{0.214} \\
	
%
     \bottomrule
     \end{tabular}
\end{table*}

\begin{figure}[t]
  \centering
  \includegraphics[width=6cm,height=5cm]{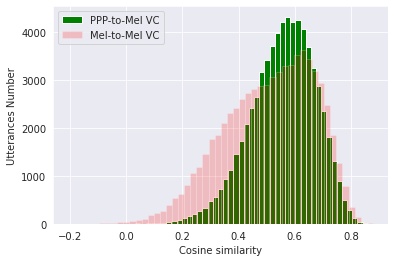}  
   \caption{Histogram of cosine similarity score on the in-set experiment.}
  \label{fig:hist}
\end{figure}

\subsubsection{Data augmentation of the Mel-to-Mel VC system }

For training the Mel-to-Mel VC system, the loss function of the many-to-many voice conversion model is 
\renewcommand{\baselinestretch}{1.0}
\begin{equation}
\begin{split}
\mathcal L_{total}&= \mathcal L_{mel\_before}+\mathcal L_{mel\_after}+\mathcal L_{stop\_token\_loss} \\
			& +5*\mathcal L_{embedding\_loss}+ \mathcal L_{regular\_loss}
\end{split}
\end{equation}


The symbolic denotes of loss function are also described in detail in\cite{tacotron2}. To make the embedding of the voice generated by the voice conversion model close to the speaker embedding, we increased the weight of embedding loss and set it to 5.

After that, we generated 200 utterances for each target speaker based on a trained Mel-to-Mel VC system. For every target speaker, the source speech of VC's input was random chosen from the other 339 speaker utterances. The embeddings generated by the VC system were computed the cosine similarity to handle the outlier. The data with similarity greater than 0.6 are retained.

%

The limited text-dependent training data (3600 utts)  were adopted as source speech for the out of set unseen speaker augmentation. The 20 utterances of each out of set unseen speakers were randomly chosen as target speaker data. Therefore, each out of set unseen speaker has 20 generated text-dependent utterances. After that, the generated data with cosine similarity less than 0.3 are filtered out. Since the out-set voice conversion is a challenging task, the threshold is not very strict (the most out of set embedding similarity is less than 0.5).

\subsubsection{Data augmentation of the PPP-to-Mel VC system }

The handing generated data processing of the PPP-to-Mel VC augmentation method is the same as the Mel-to-Mel VC system, and the loss function is designed based on \cite{tacotron2}. Different from the Mel-to-Mel VC system, since the PPP feature includes the speaker information and the target speaker PPP feature is adopted to map the target Mel-spectrogram feature in the training stage, the PPP-to-Mel VC system can not handle the out of set speaker augmentation. In addition, the result of the out of set speaker embedding similarity matrix we computed also proves the inference. Therefore, the system has a good performance in seen speakers but poor results in unseen speakers.

\begin{table}[btp]   
	\caption{The WER$[\%]$ and cosine similarity for different system on the in-set experiment.} 
	\label{tab:wer}
	\center    
	\begin{tabular}{lcll}    
	\toprule   \multirow{2}*{Model}  & Cosine/Utt Num. & Utt Num. & \multirow{2}*{WER$[\%]$} \\   
	& (average/all) &($>0.6$) & \\ 
	\midrule   
			PPP-to-Mel  & 0.555/68000 & 26029 & 9.11\\
			Mel-to-Mel  & 0.510/68000 & 23100 & 10.28\\     
	\bottomrule 
\end{tabular}  
\end{table}

For the seen speaker augmentation scenario, the word error rate (WER) and Cosine similarity are adopted as objective metrics to measure the VC systems. Fig. \ref{fig:hist} and Table.\ref{tab:wer} shows the performance of different VC system in seen speaker data. Each VC system both generated 68000 text-dependent utterances. Comparing with the Mel-to-Mel system, the PPP-to-Mel system's mean of all embedding for cosine similarity is higher. Moreover, the WER of the PPP-to-Mel VC system is less than Mel-to-Mel in retained utterances data. Therefore, the speech quality of the PPP-to-Mel system is better in objective metrics.

\subsection{Speaker augmentation based on speed perturbation}

 We use speed perturbation based on the SoX  $speed$ function that modifies the pitch and tempo of speech by resampling. The strategy also has a successful application in speech and speaker recognition tasks \cite{speed_perturb_spk,speed_perturb_speech}. The limited text-dependent speaker data created two versions of the original signal with speed factors of 0.9 and 1.1. The new classifier labels are generated at the same time when the speed perturbation creates the additional examples.
 
\section{Experiments}

Table.\ref{tab:results} shows the results of different data augmentation strategies. The performance metrics are equal error rate (EER) and minimum detection cost function (mDCF) with $P_{\textrm{target}}=0.1$. The training data of the baseline system is \texttt{9 utt}(limited text-dependent data). Since the speaker data is much less to make the model overfit, the system performance is descended seriously. On the other hand, since the pitch shift AUG augment the speaker labels, the EER of the system has been improved by nearly 10\%. The VC AUG with the PPP-to-Mel system also promotes 20\% in the term of EER. Moreover, it is easy to detect that the system combined use of pitch shift AUG and VC AUG achieved the best performance. Experimental results show that, in the scenario of limited training data, the proposed method significantly improves the EER  performances from 6.51\% to 4.51\%, and the performance of the mDCF$_{0.1}$ improves from 0.265 to 0.214.

Furthermore, since the speech quality and similarity generated by the PPP-to-Mel VC system are better than the Mel-to-Mel system, a better result is achieved by the PPP-to-Mel VC system. Nevertheless, the Mel-to-Mel VC system still explored the out of set unseen speaker augmentation and achieved a little improvement. The results obtained show that the VC Aug$_{in}$ method is reasonable, and the VC Aug$_{out}$ method also has a big margin for improvement. Based on our experiments, we can infer that the voice conversion data augmentation is reasonable. 

\section{Conclusion}

This paper proposes two data augmentation methods to improves the text-dependent speaker verification system's performance under the limited training data scenario. The results show that this strategy is feasible and effective. In the future works, we will further explore the methods and strategies for voice conversion based data augmentation with unseen or even artificiality created speaker.

\bibliographystyle{IEEEbib}
\bibliography{strings,refs,arxiv}

\end{document}